\documentclass[aps,prl,preprint,groupedaddress]{revtex4}
\usepackage{graphicx}

\begin{document}

\title{Comment on "Crossover of collective modes and positive
sound dispersion in supercritical state"
}

\author{T. Bryk$^{1,2}$, I. Mryglod$^{1}$
}

\affiliation{
$^1$Institute for Condensed Matter Physics of the
National Academy of Sciences of Ukraine,
1 Svientsitskii Street, UA-79011 Lviv, Ukraine\\
$^2$ Institute of Applied Mathematics and
Fundamental Sciences, Lviv Polytechnic National University, UA-79013
Lviv, Ukraine\\
}

\date{\today}

\begin{abstract}
We comment on an expression for positive sound dispersion (PSD) in fluids and analysis
of PSD from molecular dynamics simulations reported in the Letter by 
Fomin {\it et al} (J.Phys.:Condens.Matt. v.28, 43LT01, 2016)
\end{abstract}

\pacs{61.20.Ja,61.20.Lc, 62.60.+v}
\keywords{generalized hydrodynamics, collective excitations, dynamics of fluids,
          positive dispersion, molecular dynamics simulations}

\maketitle

In a recent Letter\cite{Fom16} the authors promoted their own idea of a "Frenkel line",
which has to separate liquid-like and gas-like fluids in supercritical region, and took
as a case study the behaviour of positive sound dispersion in fluids. The same group 
of authors in fact extended a sequence of their exotic claims, which contains: a "damping
 myth" \cite{Bra15,Bra15b}, a model of non-damped excitations applied 
to acoustic modes and non-hydrodynamic shear waves in fluids\cite{Bol12}, a 
"liquid-gas transition in the supercritical region" along a line which crosses 
the coexistence line below (!) critical temperature \cite{Bra13}, etc. 
As a matter of fact 
there is no rigorous theoretical basis for the concept of the "Frenkel line" starting from 
the basic principles of statistical theory of many-particle systems. Therefore the theory 
has been simply replaced by speculations, and a good example of these speculations is
right the Letter \cite{Fom16}, in which 
the authors claimed that they derived an expression
for positive sound dispersion (PSD) in fluids. The observed positive sound 
dispersion, which is a consequence of viscoelasticity and damping effects in liquds, 
was studied for many years and at least three explanations exist for this phenomenon on the 
basis of mode-coupling theory
\cite{Ern75}, of memory function (MF) approch \cite{Sco00} and of generalized collective 
modes (GCM)\cite{Bry10}.  In contrast to the existing theories of PSD the reported expression 
in \cite{Fom16} was not obtained within some theoretical approach, but it connects PSD with the 
existence of shear waves (transverse excitations) and 
was used to explain a dynamic crossover between liquid-like and gas-like fluids in supercritical
region. We show here that the expression for the apparent speed of sound $v_l$ in liquids 
reported in 
\cite{Fom16} 
\begin{equation}\label{vl}
v_l^2=c_s^2+4/3 v_t^2~, 
\end{equation}
where $c_s$ and $v_t$ are respectively hydrodynamic (macroscopic) speed of sound and 
apparent speed of transverse excitations,
is a consequence of several mistakes and wrong approximations. 

\noindent {\bf (i).} 
Usually in order to study PSD within some theoretical scheme one needs to obtain 
analytical expression for dispersion of longitudinal collective excitations beyond 
the hydrodynamic region as a function of wave number $\omega(k)$. Then the mechanism
of PSD as a deviation from the hydrodynamic dispersion law can be estimated.  
Fomin {\it et al}, 
however, did not try to derive any theoretical dispersion law $\omega(k)$ but started
from an expression for macroscopic (at $k\to 0$) elastic speeds of longitudinal
and transverse sound in solids. Making use of 
\begin{equation}\label{mom42}
\sqrt{\frac{B+4/3 G}{\rho}}
\end{equation}
for the high-frequency speed of longitudinal sound in liquids, 
where $B$ and $G$ are respectively bulk and shear moduli in elastic medium, 
and identifying the presence of $G$ in expression (\ref{mom42}) with the 
effect of shear waves on the longitudinal speed of sound - it is nearly the
same as to make a senseless claim that the speed of longitudinal phonons 
in crystals depends on the transverse phonons. We would like to remind, that 
for the case of non-molecular liquids all the static and time-dependent 
cross-correlations between longitudinal
and transverse currents as well as between the corresponding components of 
stress tensor are equal to zero;\\
\noindent {\bf (ii).} The {\it apparent} speed of sound in liquids estimated 
either from scattering experiments or computer simulations is not (\ref{mom42})
as it is stated in \cite{Fom16}.
In fact, all the observed peak frequencies
in experiments or simulations correspond to damped collective
excitations and one must account for damping effects. The propagation speed (\ref{mom42})
is obtained from theory in solely {\it elastic} approximation when the coupling
with density and heat fluctuations is neglected \cite{Bry08}, that results in dispersion
of non-damped (bare) high-frequency excitations, which goes at higher frequencies than the
observed ones in MD simulations or in scattering experiments;\\
\noindent {\bf (iii).} In order to modify the high-frequency dispersion with the propagation
speed (\ref{mom42}) and  
to account for the hydrodynamic dispersion a trick was suggested in \cite{Fom16}, which consists in 
replacement of the high-frequency bulk modulus $B_{\infty}$ of the elastic regime by the 
macroscopic zero-frequency one $B_0$ of the viscous regime.  Such an approximation does not 
have any sense - the high-frequency bulk modulus  $B_{\infty}$
is defined via the microscopic forces acting on particles \cite{Boo}
\begin{equation}\label{B_inf}
B_{\infty}=\frac{2nk_BT}{3}+P+\frac{2\pi n^2}{9}
\int_0^{\infty} dr g(r)r^3\frac{d}{dr}[r\frac{dV(r)}{dr}]~,
\end{equation}
while $B_0$ is connected solely with 
the macroscopic isothermal compressibility of the system 
\begin{equation}\label{B0}
B_{0}=\frac{nk_BT}{S(k\to 0)}\equiv \frac{nk_BT}{1+4\pi n\int_0^{\infty} dr [g(r)-1]r^2}~,
\end{equation}
i.e. it is impossible to reduce the expression for
$B_{\infty}$ \cite{Boo} to $B_0$ by any reasonable approximation. In (\ref{B_inf})  $n$, $k_B$,
$T$, $P$, $g(r)$, $V(r)$ are respectively the number density, Boltzmann constant, temperature,
pressure, pair distribution function and effective interatomic pair potential;\\
\noindent {\bf (iv).} Only one of the two high-frequency quantities in (\ref{mom42}) was 
replaced by its zero-frequency macroscopic value, i.e. only the bulk modulus in elastic regime
was substituted by macroscopic bulk modulus of the viscous regime, while the shear modulus 
in (\ref{mom42}) remained in the elastic regime. In order to be consistent Fomin {\it et al}
had to apply the same approximation both to $B_{\infty}$ and $G_{\infty}$. Immediately 
they would see the validity of the approximation they were using, because $G_0\equiv 0$ for 
liquids and the resulting expression will be senseless;\\
\noindent {\bf (v).} In order to have a reference speed of sound, which for estimation of the 
positive sound dispersion must be the macroscopic adiabatic speed of sound $c_s$, an expression
\begin{equation}\label{cs}
c_s=\sqrt{\frac{B_0}{\rho}}
\end{equation}
was used in \cite{Fom16}, in which the ratio of specific heats $\gamma$ is missing in numerator 
under the square root, that results in isothermal and not in adiabatic speed of sound, i.e. absolutely 
other reference speed of sound which cannot be used for estimation of PSD. In case 
the authors of \cite{Fom16} may argue that the zero-frequency bulk modulus obtained from 
the proposed substitution $B(\omega)=B_0$ is the adiabatic one, then  
this is nothing else but a misleading manipulation with quantities made in \cite{Fom16}, because 
it is impossible to reduce a high-frequency {\it elastic} quantity to the {\it adiabatic} one. 
We would like to 
remind the readers, that the adiabatic propagation is the consequence of the coupling of the isothermal 
macroscopic propagation to heat fluctuations - it cannot be obtained by any approximation from the 
high-frequency bulk modulus. And the ratio of specific heats $\gamma$ is right a measure of 
coupling between the heat fluctuations and viscous processes;\\
\noindent {\bf (vi).} The reported in \cite{Fom16} expression for the apparent speed of sound (\ref{vl})
implies that the positive sound dispersion will be observed only for wave numbers for which the shear 
waves can propagate in liquid. This immediately leads to very strange dispersion of collective 
excitations which contains 
a kink at $k_{sw}$, where $k_{sw}$ is the width of the propgation gap for shear waves, i.e. the 
smallest wave number at which shear waves (transverse excitations) can be observed. For $k<k_{sw}$
the dispersion of longitudinal excitations according to (\ref{vl}) is solely hydrodynamic one, while 
at $k_{sw}$ the dispersion changes its slope, that has never been observed either from scattering 
experiments or from computer simulations.  A simple example can be the dispersion of
collective excitations in the simplest fluids of hard spheres or 2D hard disks. For 2D hard
disk fluids in a wide range of packing fractions the non-hydrodynamic trasverse excitations
were observed\cite{Hue15} while there was no evidence found for positive sound dispersion -
only "negative" dispersion with respect to the hydrodynamic dispersion law was observed. This
fact contradicts the suggested by Fomin {\it et al} relation between the positive sound dispersion
and transverse excitations.

In their Fig.2 Fomin {\it et al} reported dispersion curves for three densities of Lennard-Jones 
fluid, obtained from peak positions of longitudinal (L) and transverse (T) current spectral 
functions $C^{L/T}(k,\omega)$ derived from molecular dynamics (MD) simulations. In order to be 
consistent
with their "Frenkel line" idea of dynamic crossover they claim that above the "Frenkel line" (for 
lower densities, open stars in their Fig.1) the positive sound dispersion and transverse excitations
disappear, that is clearly not the case as one can clearly observe in their Fig.2. Moreover, 
the excess of the apparent speed of sound over the hydrodynamic one (very similar positive sound 
dispersion as in Figs.2a and 2b) namely for low-density fluids was reported in \cite{Gor13} from 
MD simulations of supercritical Ar at four different temperatures, i.e. for about 30 thermodynamic
points of low-density Ar. Therefore, it is not clear why Fomin {\it et al} claim that there is no 
PSD for low-density states in contradiction to their Figs.2a and 2b and to the results of \cite{Gor13}. 
Furthermore, in their Fig.2c the dispersion curves were reported for 12 wave numbers, proportional 
to some minimal value less than $0.1$\AA$^{-1}$. This is in contradiction to the reported in Table~1 
value $k_{min}=0.11887$\AA$^{-1}$ that implies wrong sampling of wave numbers, which are not 
consistent with the periodicity of MD box. 

Another issue is connected with a comparison of "direct" numerical estimation of dispersion 
and GCM results.
The used in \cite{Fom16} "direct" estimation of the dispersion curves from peak positions of 
$C^{L/T}(k,\omega)$ is the most 
simple and not really precise one. In fact, the shape of $C^{L/T}(k,\omega)$ in that "direct" approach 
is solely ascribed to respectively L/T collective excitations and neglects contributions coming
from other propagating and relaxing processes. More precise approaches take 
into account contributions from different relaxation processes and coupling of L and T currents 
with them - one uses 
either a fit within the memory function approach \cite{Sco00}, fit-free GCM approach \cite{Mry95,
Bry01,Bry10}, or a combination of a fit with GCM methodology\cite{Bry13,Bry16}. The GCM 
approach \cite{Key71,Kiv72,deS88,Mry98} is able to separate contributions from different collective 
relaxing and propagating modes and simultaneously satisfies any required level of exact sum rules. 
The quality of the GCM approach for description of time correlation functions and corresponding 
spectral functions was shown many times \cite{Bry01b,Bry13,Bry16} - therefore it is strange that 
Fomin {\it et al} claim that the numerical estimation of dispersion from peak positions of 
$C^{L/T}(k,\omega)$ is superior to the GCM approach. It is known for long time that the stronger
the contribution from relaxation processes (like for low-density fluids with essentially 
different from unity ratio of specific heats $\gamma$, or for dense fluids
in the region of de~Gennes slowing down of density fluctuaions) the larger will be the deviation of 
the observed peak position of $C^{L}(k,\omega)$ from the real frequency of collective excitation. 
It seems Fomin {\it et al} do not realise that the L-dispersion curves estimated from the observed 
peaks of MD-derived either dynamic structure factors $S(k,\omega)$, or
imaginary part of dynamic susceptibility $\chi^{''}(k,\omega)$, or L-current spectral function 
$C^{L}(k,\omega)$ will be different, while the corresponding frequencies from the GCM analysis of
these three spectral functions will be identical as it must be.   Concerning the other strange claims 
of Fomin {\it et al} on the choice of dynamic variables and sum rules in GCM approach  
we refer the readers to our extended Comment\cite{Bry15}.

\noindent


%
%
\end{document}